\documentclass[conference]{IEEEtran}
\IEEEoverridecommandlockouts

\usepackage{cite}
\usepackage{amsmath,amssymb,amsfonts}
\usepackage{algorithmic}
\usepackage{graphicx}
\usepackage{textcomp}
\usepackage{xcolor}
\usepackage[hidelinks]{hyperref}
\usepackage{booktabs}
\usepackage{multirow}
\usepackage{subcaption}
\usepackage{dblfloatfix}
\usepackage{placeins}
\usepackage{siunitx}
\usepackage{orcidlink}

\makeatletter
\let\IEEE@origfigure\figure
\let\IEEE@origendfigure\endfigure
\renewenvironment{figure}[1][\fps@figure]{%
  \IEEE@origfigure[#1]%
  
}{\IEEE@origendfigure}
\let\IEEE@origfigurestar\figure
\let\IEEE@origendfigurestar\endfigure
\renewenvironment{figure*}[1][\fps@figure]{%
  \IEEE@origfigurestar[#1]%
  
}{\IEEE@origendfigurestar}
\makeatother

\def\BibTeX{{\rm B\kern-.05em{\sc i\kern-.025em b}\kern-.08em
    T\kern-.1667em\lower.7ex\hbox{E}\kern-.125emX}}

\begin{document}
\captionsetup{font=small}
\captionsetup[sub]{font=small}

\author{
  \IEEEauthorblockN{
    Maickol Fernandez-Obando\orcidlink{0009-0001-8132-9312},
    Luis G. Leon-Vega\orcidlink{0000-0002-3263-7853},
    Leonardo Cardinale-Villalobos\orcidlink{0000-0002-9649-6017},\\
    Christopher Vega-Sanchez\orcidlink{0000-0002-2174-8291},
    Luis D. Murillo-Soto\orcidlink{000-0002-6601-1082}\\
  }
  \vspace{3pt}
  \IEEEauthorblockA{
      Costa Rica Institute of Technology, Cartago 159-7050, Costa Rica\\
  }
  \vspace{4pt}
  \IEEEauthorblockA{
    \begin{tabular}{ccc}
      {l.leon,lcardinale}@tec.ac.cr
    \end{tabular}
  }
}

\title{Design and Implementation of a Multi-Sensor DAQ System for Comparative Photovoltaic Performance Analysis}

\maketitle

\begin{abstract}
The rigorous analysis of specialized physical processes often demands custom data acquisition architectures that offer flexibility and precision beyond the capabilities of general-purpose commercial loggers. This paper presents the design and implementation of a robust data acquisition system (DAQ) for a comparative analysis of the performance of two photovoltaic panels with two different cooling systems. The system integrates a custom PCB design for 20 thermistors, dual high-precision INA228 current/voltage sensors, environmental monitoring equipment, and a Raspberry Pi 4-based acquisition platform. The software architecture implements autonomous operation with enhanced fault recovery, dual storage redundancy (local CSV and InfluxDB), cloud synchronization via Google Drive, and real-time visualization through Grafana dashboards. Field deployment demonstrated system reliability, including automatic recovery from power interruptions, a 1-minute sampling rate, remote monitoring capabilities, and continuous operation during a 5 AM to 6 PM daily window. The modular hardware and software architecture enables simultaneous monitoring of two photovoltaic panels for research on direct performance comparison under identical environmental conditions.
\end{abstract}


\section{Introduction}

Our planet is experiencing a warming phase caused by greenhouse gasses, which causes an increase in global temperature. High ambient temperatures and heat waves significantly impact the performance of electricity production from photovoltaic (PV) panels installed around the world \cite{Razak_A}.

The electrical energy coming from photovoltaic systems represents more than 10\% of global energy production, and it accounts for almost 80\% of the increase in renewable energy projected for 2025\cite{IEA2025}. However, photovoltaic panels decrease their productivity when they heat up; therefore,  temperature is a key variable to improve productivity in such systems. For this reason, research has been conducted to reduce the temperature of photovoltaic modules \cite{Sheik2022,Mahdavi_A}. However, for these developments to be reliable, it is important to closely monitor electrical, thermal, and environmental variables related to photovoltaic modules using data acquisition systems.

Building on this premise, research on the thermal behaviour of PV systems demands high-resolution data acquisition for performance characterization and comparative analysis. Flexible sensing stacks are particularly critical on research benches, where instrumentation must be reconfigured frequently, and thermal gradients must be mapped across paired panels to expose treatment deltas. 
Conventional PV loggers and supervisory control platforms emphasize fleet-level KPIs, aggregated energy reporting, and utility-class safety limits. Those priorities leave little room for dense thermal imaging, per-module instrumentation, or rapid sensor reconfiguration, all of which are necessary when comparing experimental cooling strategies against a reference module under identical conditions.

\begin{figure}[htbp]
\centering
\includegraphics[width=\columnwidth]{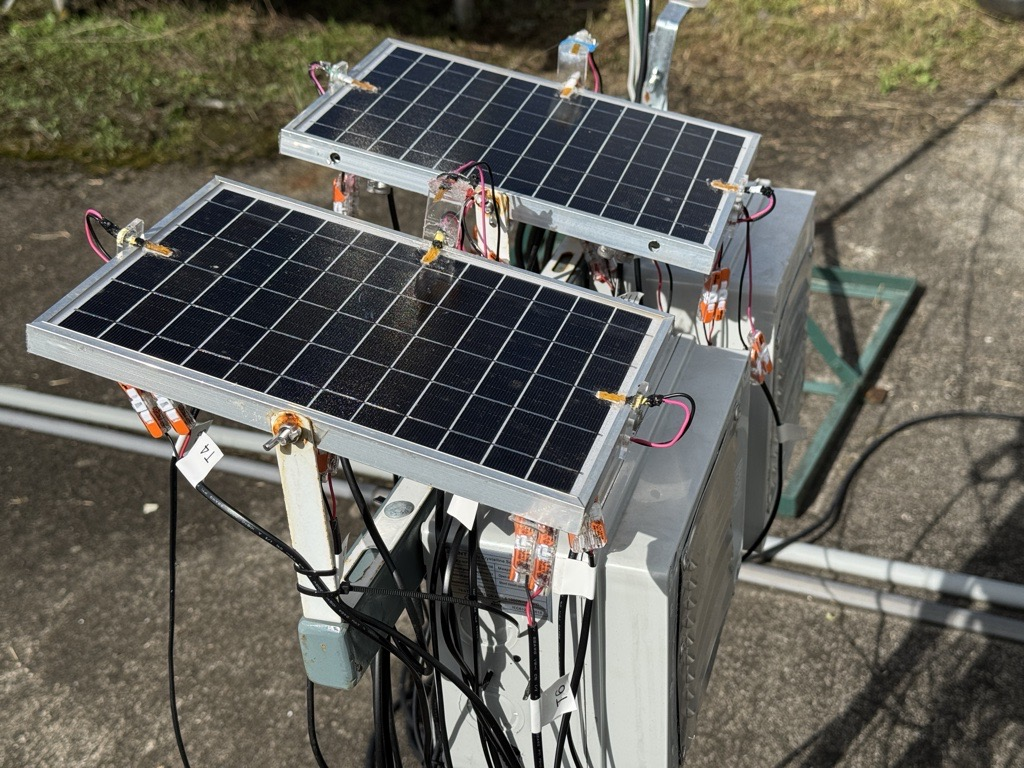}
\caption{PV modules used to evaluate different cooling systems aimed at reducing PV average temperture, requiring the implementation of the DAQ proposed in this research.}
\label{fig:PV_modules}
\end{figure}

In this regard, this research presents a custom-designed data acquisition system for a passive microfluidic heat exchanger, which dissipates heat without requiring external energy sources. This low-cost, modular, portable, and easy-to-install heat exchanger could be installed on the millions of photovoltaic panels installed worldwide, for both mono- and polycrystalline monofacial technologies, to increase the efficiency of the panels and, consequently, the efficiency of solar farms. Figure~\ref{fig:PV_modules} shows the system used to carry out this research. It consists of two 6 W monocrystalline PV modules: a control panel and a passive-cooled panel equipped with a heat exchanger. Given that both modules are placed next to each other, it can be considered that they operate under identical environmental conditions but using independent MPPT controllers.

The proposed DAQ solution presented here combines a large number of thermal sensors, independent electrical channels, and a data infrastructure to compare the two identical systems under different cooling systems. The main established requirements were: (i) spatial measurements of temperature and environmental variables, (ii) collection of the main electrical variables, and (iii) automated data acquisition adapted to actual operating conditions.

This work contributes to the design and implementation of a Raspberry Pi-based multi-sensor DAQ system for temperature and power analysis oriented to cooling-system research. The architecture delivers dense thermal gradient data acquisition through dense thermistor arrays per panel, dual electrical measurement chains, and integrated meteorological context while remaining cost-effective for research benches. Field results, later in the paper, demonstrate how these capabilities enable direct comparison between passive-cooled and control panels.

\noindent This manuscript is organized as follows: Section~\ref{sec:comparison} covers the analysis on commercial and academic monitoring solutions. Section~\ref{sec:system-architecture} summarizes the proposed system architecture, while Section~\ref{sec:hardware} and Section~\ref{sec:software} detail the hardware and software stacks, respectively. Section~\ref{sec:results} presents the field deployment results, and Section~\ref{sec:conclusion} presents the most important findings of this research.

\section{Comparison with Existing Solutions}\label{sec:comparison}

Photovoltaic monitoring solutions span from utility-scale commercial platforms to research-oriented acquisition systems, each addressing different operational priorities. Commercial platforms prioritize fleet management and regulatory compliance, while academic systems emphasize flexibility and cost-effectiveness, often at the expense of robustness or channel density. This section examines representative solutions across both domains to contextualize the requirements for comparative cooling systems experiments.

Commercial PV monitoring platforms are designed for utility-scale arrays rather than controlled experiments. The Solar-Log line, for instance, combines metering-certified power meters with proprietary gateways aimed at multi-string plants above tens of kilowatts and focuses on energy-yield KPIs and fleet alarms~\cite{solar_log_datasheet}. While robust for large installations, it is optimized for aggregated DC inputs and string-combiner integration, leaving no provision for attaching dense thermistor arrays or low-level meteorological probes per module. Configuration assumptions (PV strings above tens of amperes, 1~kV DC limits, external pyranometers) make them impractical for side-by-side research benches where sensors must be rewired frequently following research emerging interests.

At the opposite end of the cost-complexity spectrum, research-oriented DAQ examples also prove insufficient. The real-time DAQ study in~\cite{daq_research} uses a general-purpose microcontroller with a modest ADC front-end and a cloud dashboard. It demonstrates streaming of irradiance and electrical variables, yet relies on single-channel measurement loops with limited buffering through basic SD card storage. The architecture lacks redundancy, multi-threaded acquisition, or distributed thermal instrumentation, thereby preventing unattended comparisons between competing panel treatments.

Research-grade commercial dataloggers represent an intermediate option. The Campbell Scientific CR1000Xe measurement and control datalogger~\cite{campbell_cr1000xe} provides industrial-grade reliability with 16 single-ended analog inputs (expandable via multiplexers), 24-bit ADC resolution, extended temperature operation (\SIrange{-55}{+85}{\degreeCelsius}), and programmable sampling from \SI{1}{\milli\second} to daily intervals. Its CRBasic programming environment and multiple communication protocols (Ethernet, cellular, satellite) enable sophisticated monitoring deployments. However, at \$2,892.25 USD for the base datalogger alone~\cite{campbell_cr1000xe_price}—before sensors, enclosures, power supplies, or software licenses—making them prohibitive for budget-constrained research programs.

To ground the cost-effectiveness claims, Table~\ref{tab:cost_comparison} reports the bill of materials for the proposed DAQ using Amazon-sourced 2025 USD prices. The full edge node totals \$331, including \$50 for the PCB stack-up (fabrication at \$10 plus \$40 international shipping) and a breakdown across the main categories requested (Raspberry Pi, PCBs, integrated circuits, and sensors). This represents roughly 11\% of the CR1000Xe base price, before accounting for the latter's additional sensor and enclosure costs.

\begin{table}[htbp]
\caption{Estimated bill of materials for the DAQ (2025 USD).}
\label{tab:cost_comparison}
\centering
\begin{tabular}{l r}
\toprule
Raspberry Pi 4 & \$65 \\
PCBs (fabrication, shipping, passives) & \$56 \\
Integrated circuits (ADC, INA228, 74HC4051) & \$52 \\
Sensors (thermistors, DHT22, WeatherKit) & \$108 \\
Power and protection (CN3791, batteries, ceramic load) & \$50 \\
\midrule
\textbf{Total} & \textbf{\$331} \\
\bottomrule
\end{tabular}
\end{table}

Advanced thermal characterization approaches highlight the importance of spatial temperature resolution. Li~\emph{et~al.}~\cite{li_fbg_sensors_2022} demonstrate distributed fiber Bragg grating (FBG) sensing for PV hot-spot detection, deploying 9 FBG temperature sensors across a single 960$\times$\SI{480}{\milli\metre} panel with $\pm$ \SI{0.1}{\degreeCelsius} resolution at \SI{100}{\hertz} sampling. The wavelength-multiplexed array enables continuous spatial thermal mapping validated against infrared thermography, proving that multi-point temperature monitoring directly improves diagnostic capability for performance anomalies. However, the optical approach requires specialized demodulation equipment, individual sensor calibration, and careful strain-isolated installation, resulting in higher complexity and cost than contact-based thermistor methods. While the 9-sensors FBG configuration for temperature measurement represents the state-of-art in non-contact thermal monitoring, achieving higher spatial resolution with simpler instrumentation remains an open challenge for research applications.

Modern IoT architectures demonstrate the viability of open-source monitoring stacks. Silva~\emph{et~al.}~\cite{silva_iot_microgrid_2022} implement a Raspberry~Pi-based supervisory system with InfluxDB time-series storage and Grafana dashboards for microgrid monitoring, validating this platform as "more fault-resilient" and "easily expandable" than commercial SCADA alternatives. Their three-tier architecture (field sensors $\rightarrow$ InfluxDB $\rightarrow$ Grafana) operates entirely on local edge hardware with configurable polling rates and multi-user access. However, the system lacks external backup redundancy—SD~card failure or power loss could result in permanent data loss—and focuses on power-flow monitoring rather than high-density thermal characterization. The platform does not address analog multiplexing for large thermistor arrays or specialized signal conditioning for low-level temperature measurements essential for comparative cooling studies.

These gaps motivate the design of a more complete solution that combines (i)~dense spatial thermal sensing exceeding optical FBG arrays in sensor count, (ii)~proven open-source IoT architecture enhanced with dual storage redundancy and cloud synchronization, (iii)~custom analog multiplexing hardware enabling cost-effective high-channel-count acquisition, and (iv)~specialized fault recovery mechanisms for unattended long-term field deployment. The following sections detail how custom PCB design, multi-threaded software architecture, and experimental-focused sampling strategies address these requirements while maintaining research-accessible cost and complexity.

\section{System Architecture}\label{sec:system-architecture}

The proposed architecture brings together sensing, acquisition, storage, and visualization layers to support paired-panel experiments. Each PV module—one with the cooling treatment and one reference—hosts ten thermistors routed through a multiplexed PCB, dual INA228 channels for electrical telemetry~\cite{ina228_datasheet}, and shared meteorological probes. A Raspberry Pi 4~\cite{raspberrypi:documentation} coordinates these subsystems, polling sensors at one-minute intervals, buffering locally, and forwarding summaries to InfluxDB~\cite{influxdb_docs} and cloud storage so comparisons remain available even after field interruptions. Fig.~\ref{fig:architecture} outlines the resulting data flow from sensors to dashboards.

\subsection{Functional Overview}

\begin{figure}[htbp]
\centering
\includegraphics[width=\columnwidth]{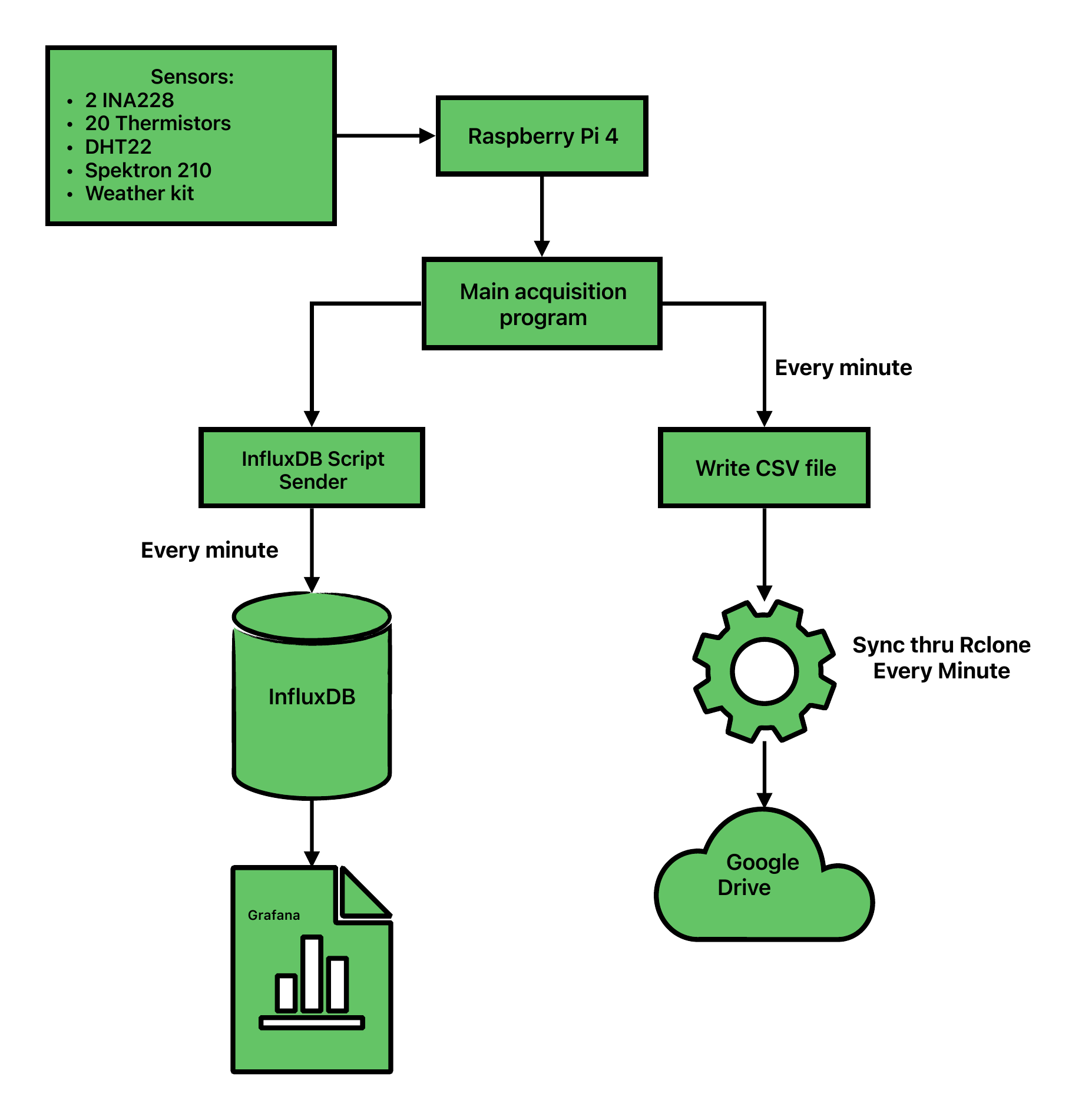}
\caption{System architecture showing data flow from sensors through acquisition, storage, and visualization layers.}
\label{fig:architecture}
\end{figure}

The system architecture follows a multi-layered design separating hardware interfacing, data acquisition, storage, and presentation (Fig.~\ref{fig:architecture}), composed of two different systems: one at the edge and another at the cloud.

At the edge, a Raspbery Pi 4 is in charge of the hardware interfacing, data acquisition, and redundant local storage. It runs a Python-based DAQ software that performs sensor polling, getting data from environment and temperature sensors at different rates, particularly retrieving meteorological and environmental information, electrical, and temperature measurements. The DAQ is also in charge of preprocessing the data, converting them to the proper scales and units, for later storage into the local non-volatile memory (through CSV files) and transfer to the cloud (to InfluxDB).

In the cloud, this work proposes a microservice-based architecture, composed of two microservices: (i) InfluxDB for time-series cloud storage, and (ii) Grafana for data visualization. Both are open-source alternatives for time-series databases and data visualization, offering a flexible alternative against commercial alternatives, and adopted by similar work~\cite{influxdb_grafana_1,influxdb_grafana_2}. Moreover, for redundancy, the local CSV files are uploaded frequently to a cloud storage using Rclone utility~\cite{rclone_docs}.


\subsection{Sensor Integration}

Comparing altered and reference panels requires three complementary sensing fronts: precise electrical telemetry to quantify delivered power, dense thermal readings to observe how cooling strategies redistribute temperature, and environmental context to decouple weather-driven effects from treatment responses. Accordingly, the system integrates the following sensor array:

\begin{itemize}
\item \textbf{Electrical:} Dual INA228 power monitors~\cite{ina228_datasheet} (I2C addresses 0x40, 0x41) measuring voltage, current, power, and accumulated energy for each PV panel with 20-bit ADC resolution, capturing the subtle yield differences introduced by the cooling treatment.

\item \textbf{Thermal:} 20 NTC thermistors (10 per module) interfaced through analog multiplexers, providing spatial temperature distribution across panel surfaces to identify thermal gradients, and the effective footprint of the cooling intervention. The thermistors utilize a voltage divider configuration for signal conditioning.

\item \textbf{Environmental:} DHT22 temperature/humidity sensor~\cite{dht22_datasheet}, Spektron 210 differential irradiance sensor~\cite{tritec_spektron_210}, cup anemometer, wind vane, and tipping-bucket rain gauge~\cite{fine_offset_weather_datasheet} forming a complete meteorological station that captures the boundary conditions needed for a complete energy and thermal analysis.
\end{itemize}

The data acquisition timing and processing logic vary by sensor type to accommodate different signal dynamics. Table \ref{tab:daq_sampling} summarizes the sampling rates and processing methods for each variable. All the variables were archived at one minute interval.


\begin{table}[htbp]
    \centering
    \caption{Sampling rates and processing methods defined for each variable in the DAQ.}
    \label{tab:daq_sampling}
    \begin{tabular}{@{}lcc@{}}
        \toprule
        \textbf{Sensor / Variable} & \textbf{Sampling ($T_s$)} & \textbf{Processing Method} \\ \midrule
        Thermistor Banks           & 5 s                       & Rolling Avg (1 min)        \\
        DHT22 (Ambient Hum/Temp)           & 5 s                       & Rolling Avg (1 min)        \\
        Wind Speed                 & 1 s (1 Hz)                & Average (60 s)             \\
        Rainfall                   & Event-based               & Accumulation               \\
        Irradiance                 & 60 s                      & Instantaneous              \\
        Wind Vane                  & 60 s                      & Instantaneous              \\
        INA228 (Power)             & 60 s                      & Instantaneous              \\ \bottomrule
    \end{tabular}
\end{table}

This sensor configuration enables correlation analysis of environmental conditions, thermal distribution, and electrical performance for both control and treatment PV panels under identical ambient conditions, supporting further research on thermal cooling systems for PV cells.

\section{Hardware Design}\label{sec:hardware}

\subsection{Functional Architecture}

\begin{figure}[htbp]
\centering
\includegraphics[width=\columnwidth]{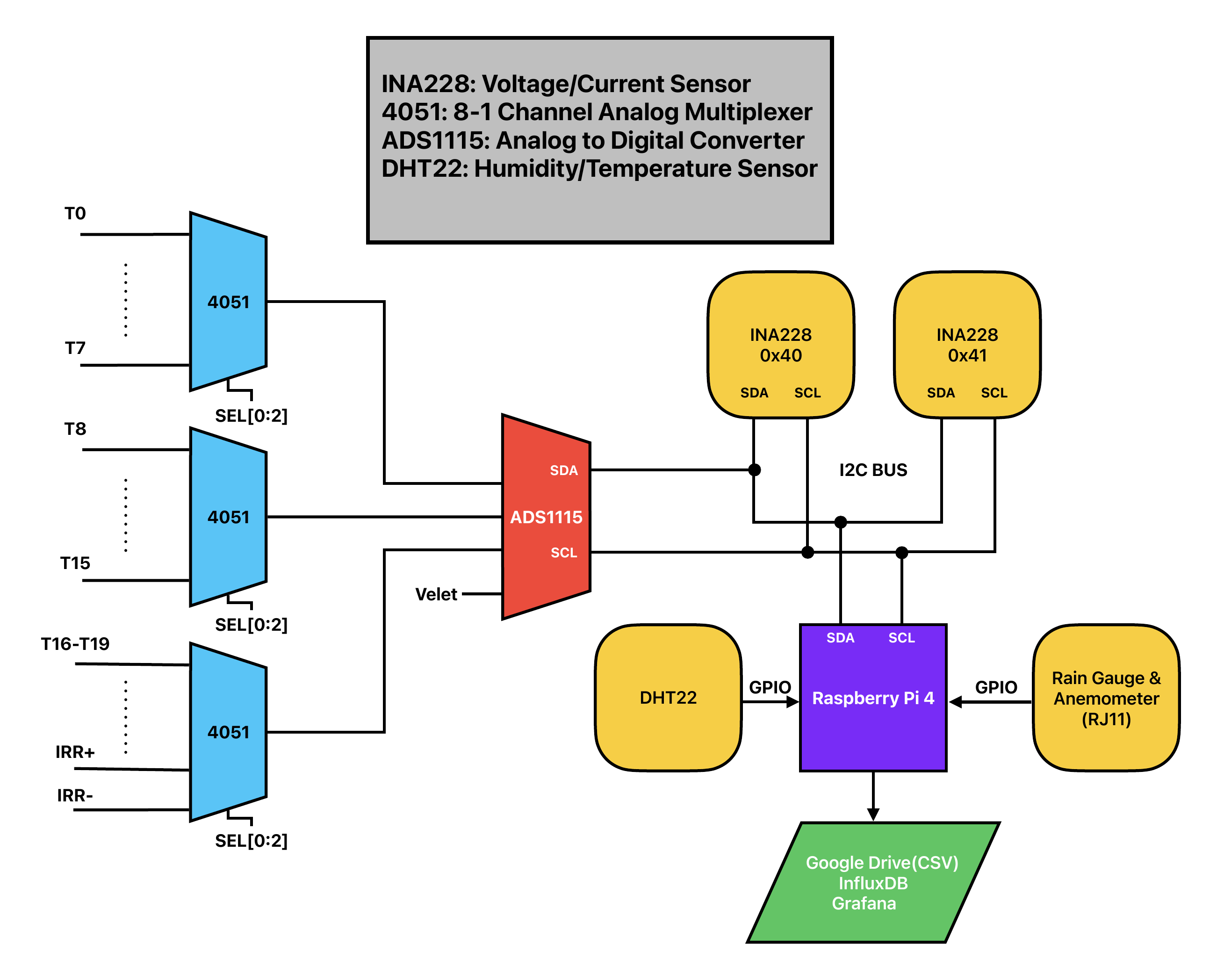}
\caption{Functional block diagram showing sensor aggregation through analog multiplexers, I2C peripherals, and GPIO interfaces converging at the Raspberry Pi 4 controller.}
\label{fig:functional}
\end{figure}

The hardware architecture implements a hierarchical sensor aggregation strategy centered on the Raspberry Pi 4~\cite{raspberrypi:documentation} platform (Fig.~\ref{fig:functional}). The design addresses the fundamental challenge of interfacing 20+ sensor channels through resource-constrained embedded hardware by leveraging analog multiplexing for thermistor arrays and I2C bus aggregation for digital sensors.

\textbf{Analog Signal Path:} The 20 NTC thermistors plus differential irradiance sensors connect through three CD74HC4051~\cite{nxp_74hc4051_datasheet} 8:1 analog multiplexers arranged in a synchronized bank. Each thermistor requires a voltage divider signal conditioning before multiplexer input, providing \SIrange{0}{3.3}{\volt} compatible with subsequent ADC stages. The Raspberry Pi controls all multiplexers simultaneously through shared select lines SEL[0:2], executing a complete scan of all thermistors every \SI{5}{\second} so that ADC conversion windows (\SI{16}{\milli\second} per channel at the configured gain) meet the bandwidth needed for minute-scale reporting. The differential irradiance sensor taps share the third multiplexer but are read during the once-per-minute reporting pass, with intentional \SI{100}{\milli\second}  settling delays between IRR$-$ and IRR$+$ to ensure adequate reading.

\textbf{Analog-to-Digital Conversion:} A single ADS1115~\cite{ti_ads1115} 16-bit I2C ADC serves as the central conversion point for all multiplexed analog signals. Four ADC channels receive three multiplexer outputs (thermistor groups and irradiance) plus a direct wind vane direction input. The 16-bit resolution delivers \SI{0.1}{\milli\volt} precision sufficient for temperature differential analysis across panel surfaces.

\textbf{Digital Power Monitoring:} Dual INA228~\cite{ina228_datasheet} high-side current/voltage monitors interface directly to each PV panel output via I2C bus. Operating at distinct addresses (0x40, 0x41), these sensors provide independent measurement of voltage, current, instantaneous power, and accumulated energy with 20-bit ADC resolution. Each INA228 includes integrated shunt resistor current sensing and programmable averaging for noise reduction; the acquisition loop polls both devices once per minute while keeping the on-chip averaging window at 1024 samples with \SI{1.052}{\milli\second}  conversion timing, yielding sub-milliamp repeatability without exceeding the system bandwidth.

\textbf{Environmental Sensors:} The DHT22~\cite{dht22_datasheet} (ambient temperature/humidity sensor), anemometer, and tipping-bucket rain gauge were connected directly via GPIO pins to minimize I2C bus load. The DHT22 employs a proprietary single-wire protocol, while the anemometer and rain gauge operate through interrupt-driven pulse counting to capture high-speed events without blocking the main processor. The wind vane was read via the ADC. This configuration ensures precise signals capture for the processing windows defined in Table~\ref{tab:daq_sampling}


\subsection{Channel Allocation Strategy}

The system implements strategic channel allocation across multiplexer and I2C address spaces to maximize throughput and minimize acquisition latency. Table~\ref{tab:mux_config} summarizes the multiplexer channel distribution. MUX1 and MUX2 each handle 8 thermistors (T0-T7 and T8-T15), fully utilizing available channels for panel thermal mapping. MUX3 combines the remaining 4 thermistors (T16-T19) with differential irradiance measurements (IRR+, IRR-), leaving two channels available.

\begin{table}[htbp]
\caption{Analog Multiplexer Channel Configuration}
\begin{center}
\small
\begin{tabular}{|c|c|c|c|}
\hline
\textbf{MUX} & \textbf{Channel} & \textbf{S2 S1 S0} & \textbf{Signal} \\
\hline
\multirow{8}{*}{MUX1} & 0 & 000 & T0 \\
 & 1 & 001 & T1 \\
 & 2 & 010 & T2 \\
 & 3 & 011 & T3 \\
 & 4 & 100 & T4 \\
 & 5 & 101 & T5 \\
 & 6 & 110 & T6 \\
 & 7 & 111 & T7 \\
\hline
\multirow{8}{*}{MUX2} & 0 & 000 & T8 \\
 & 1 & 001 & T9 \\
 & 2 & 010 & T10 \\
 & 3 & 011 & T11 \\
 & 4 & 100 & T12 \\
 & 5 & 101 & T13 \\
 & 6 & 110 & T14 \\
 & 7 & 111 & T15 \\
\hline
\multirow{6}{*}{MUX3} & 0 & 000 & T16 \\
 & 1 & 001 & T17 \\
 & 2 & 010 & T18 \\
 & 3 & 011 & T19 \\
 & 4 & 100 & IRR- \\
 & 5 & 101 & IRR+ \\
\hline
\end{tabular}
\label{tab:mux_config}
\end{center}
\end{table}

The ADS1115 ADC channels map as follows: A3 (MUX1 output), A2 (MUX2 output), A1 (MUX3 output), and A0 (direct wind direction input). This configuration requires 24 sequential ADC reads (22 multiplexed channels + 1 direct) per acquisition cycle, completed within the one-minute sampling interval.

\section{Software Architecture}\label{sec:software}

\subsection{Main Acquisition Program}
\begin{figure}[htbp]
\centering
\includegraphics[width=1\columnwidth]{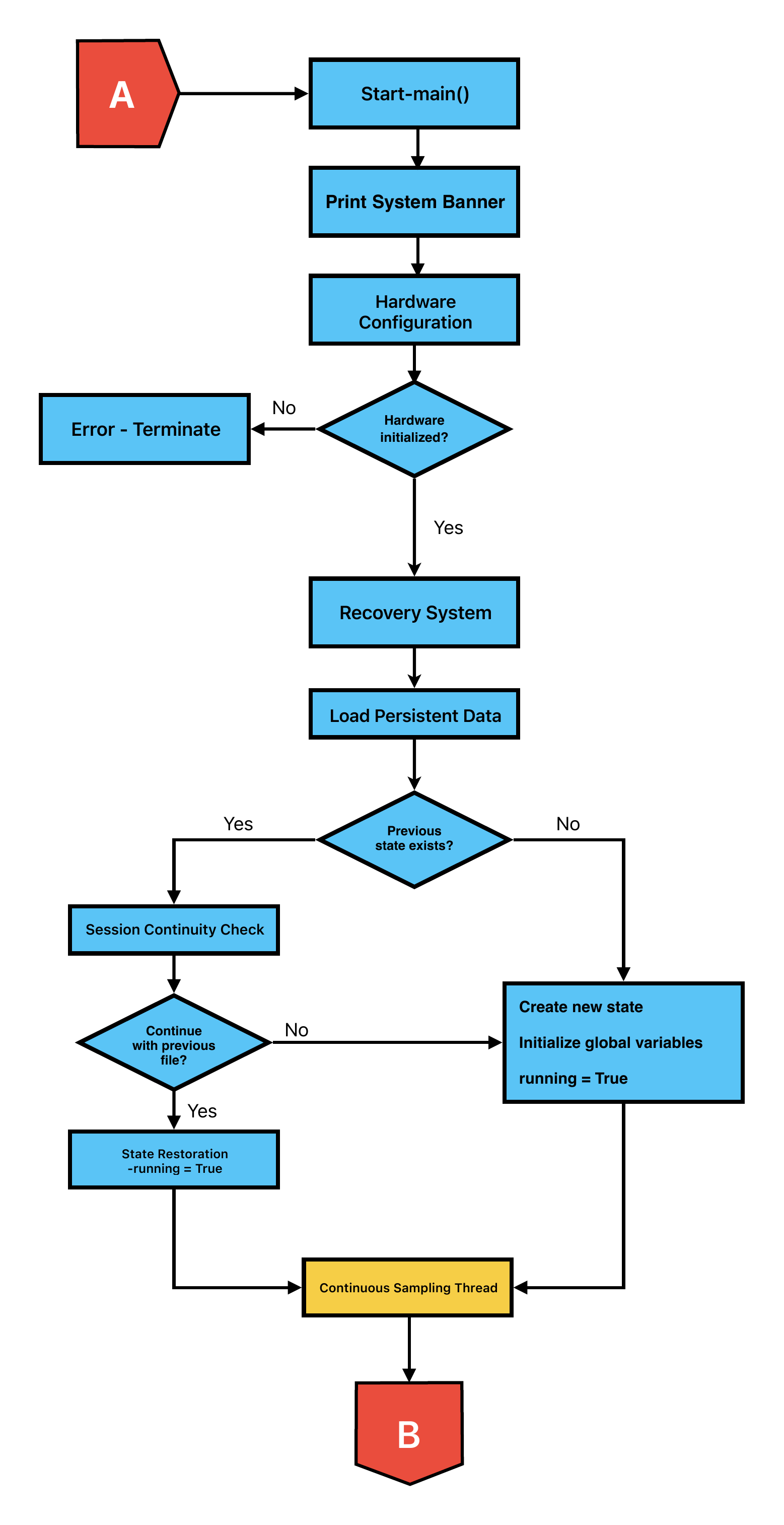}
\caption{Main program initialization flow showing hardware setup, state recovery, and main loop entry.}
\label{fig:main_flow}
\end{figure}

The local DAQ program implements a multi-threaded architecture separating fast-sampled variables (wind speed and temperature) from scheduled measurements (electrical parameters, complete system state). Fig.~\ref{fig:main_flow} illustrates the initialization and main loop structure.

\textbf{Initialization Module:} System startup begins with hardware configuration, establishing communication with all I2C devices, GPIO pins, and data interfaces. Following successful hardware initialization, the recovery subsystem attempts to restore the previous operational state by loading persistent configuration data when available.

\textbf{State Recovery Module:} The recovery logic implements intelligent session continuation based on multiple criteria: calendar day consistency, operational time window compliance (5 AM - 6 PM), data file integrity verification, and previous recording status. Successful recovery restores accumulated counters and file metadata to maintain measurement continuity across system restarts. Fresh sessions initialize new state variables and create new data files with appropriate timestamps.

\begin{figure}[htbp]
\centering
\includegraphics[width=1\columnwidth]
{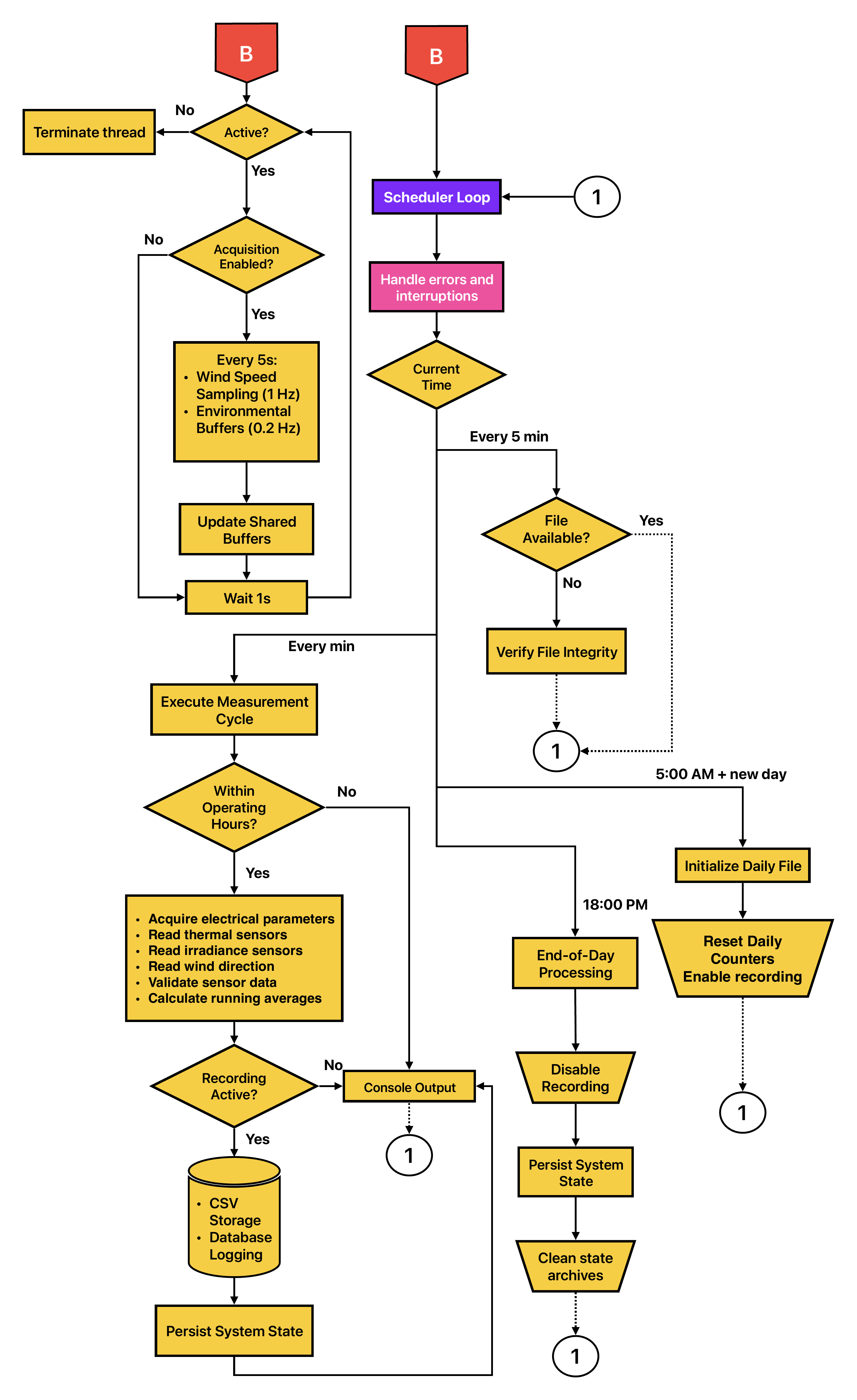}
\caption{Main loop and secondary thread operation showing scheduled tasks and continuous sampling.}
\label{fig:thread}
\end{figure}

\textbf{Multi-threaded Scheduler:} The main execution thread manages time-scheduled tasks (Fig.~\ref{fig:thread}):
\begin{itemize}
\item \textbf{5:00 AM (new day):} Daily data file creation and counter reset.
\item \textbf{Every minute:} Complete measurement cycle with database and local storage logging.
\item \textbf{Every 5 minutes:} System health verification and data integrity checks.
\item \textbf{6:00 PM:} End-of-day processing, state persistence, and resource cleanup.
\end{itemize}

A secondary sampling thread provides continuous high-frequency data acquisition: wind speed measurements at \SI{1}{\hertz} and thermistor/humidity buffer updates at \SI{0.2}{\hertz}. Thread-safe access to shared data structures employs multi-threading synchronization mechanisms, preventing race conditions during concurrent hardware bus operations.

\subsection{Error Handling and Recovery}

\begin{figure}[htbp]
\centering
\includegraphics[width=1\columnwidth]{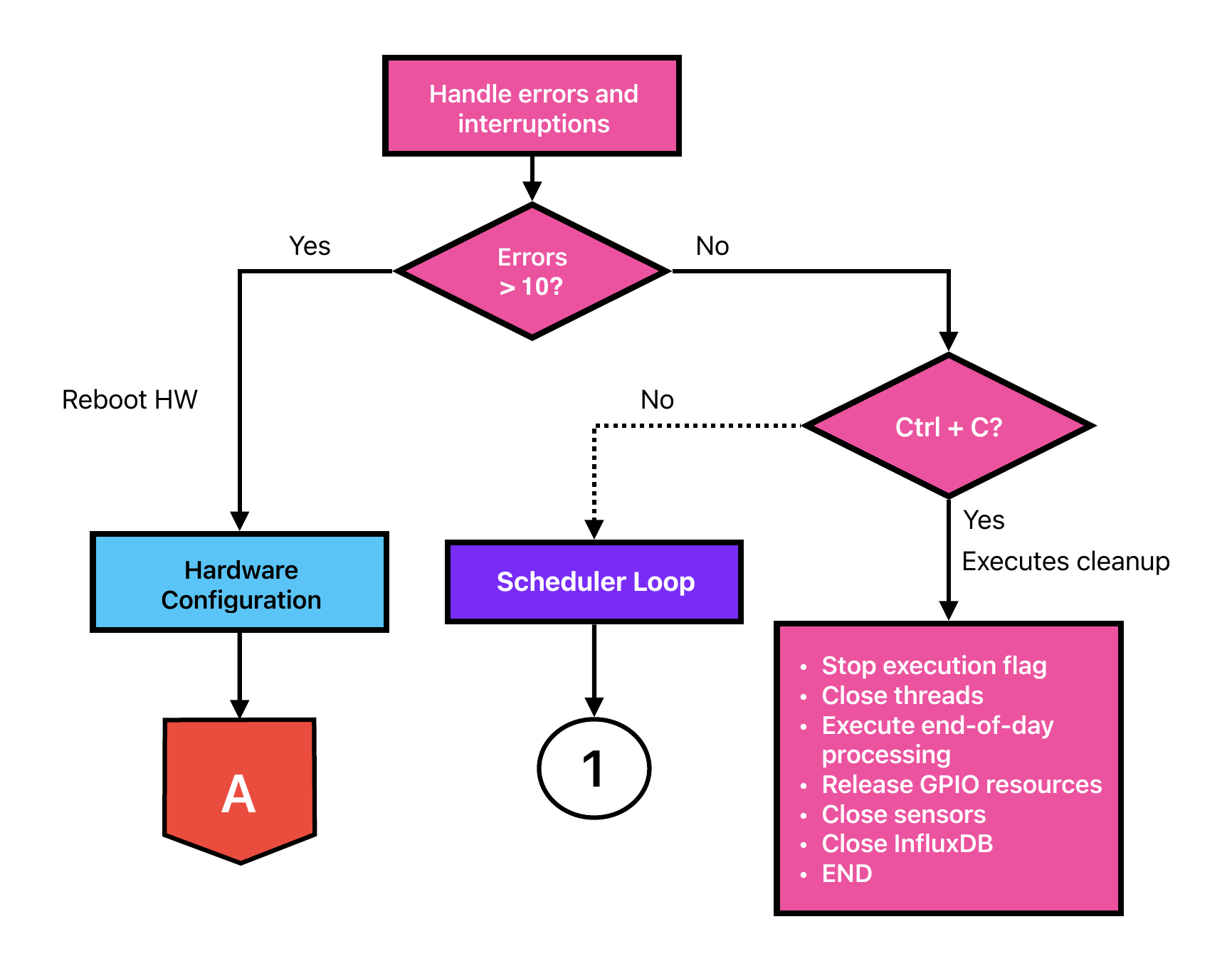}
\caption{Error-handling and cleanup procedure ensuring graceful shutdown and proper state preservation. }
\label{fig:error}
\end{figure}

The software implements robust error handling across multiple layers (Fig.~\ref{fig:error}). Individual sensor read failures log errors without terminating the program, allowing partial data capture during temporary hardware faults. Accumulated error counting triggers full hardware reinitialization after 10 consecutive failures, addressing transient I2C bus lockups or sensor resets.

 Shutdown handling (SIGINT) executes ordered clean-up: stopping acquisition threads, executing end-of-day processing, saving system state, releasing GPIO resources, and closing database connections. This procedure ensures data integrity during manual stops initiated due to process interruptions.

\subsection{Automation and Monitoring Scripts}
Bash automation scripts provide system-level orchestration without operator intervention. The boot process fixes execution permissions when necessary and spawns dedicated processes for acquisition and log monitoring. Once the primary services are running, the orchestration script is executed, verifying that both processes remain active and recording status updates to a log file; if a process terminates, the monitoring loop logs a warning without attempting automatic restart, understanding that the closure of a process will also terminate the associated Python acquisition process.

Our work also provides an interactive control for day-to-day maintenance, providing the most common administrative actions through numbered shortcuts. Users can monitor log streams, review recent entries, search errors with pattern matching, inspect log statistics, review the autostart log, and control the acquisition service. This script-centric workflow reduces SSH session time and standardizes troubleshooting steps while keeping interventions visible in the shared log files.

\section{Results and Discussion}\label{sec:results}

\subsection{Deployment Configuration}

\begin{figure}[htbp]
\centering
\includegraphics[width=\columnwidth]{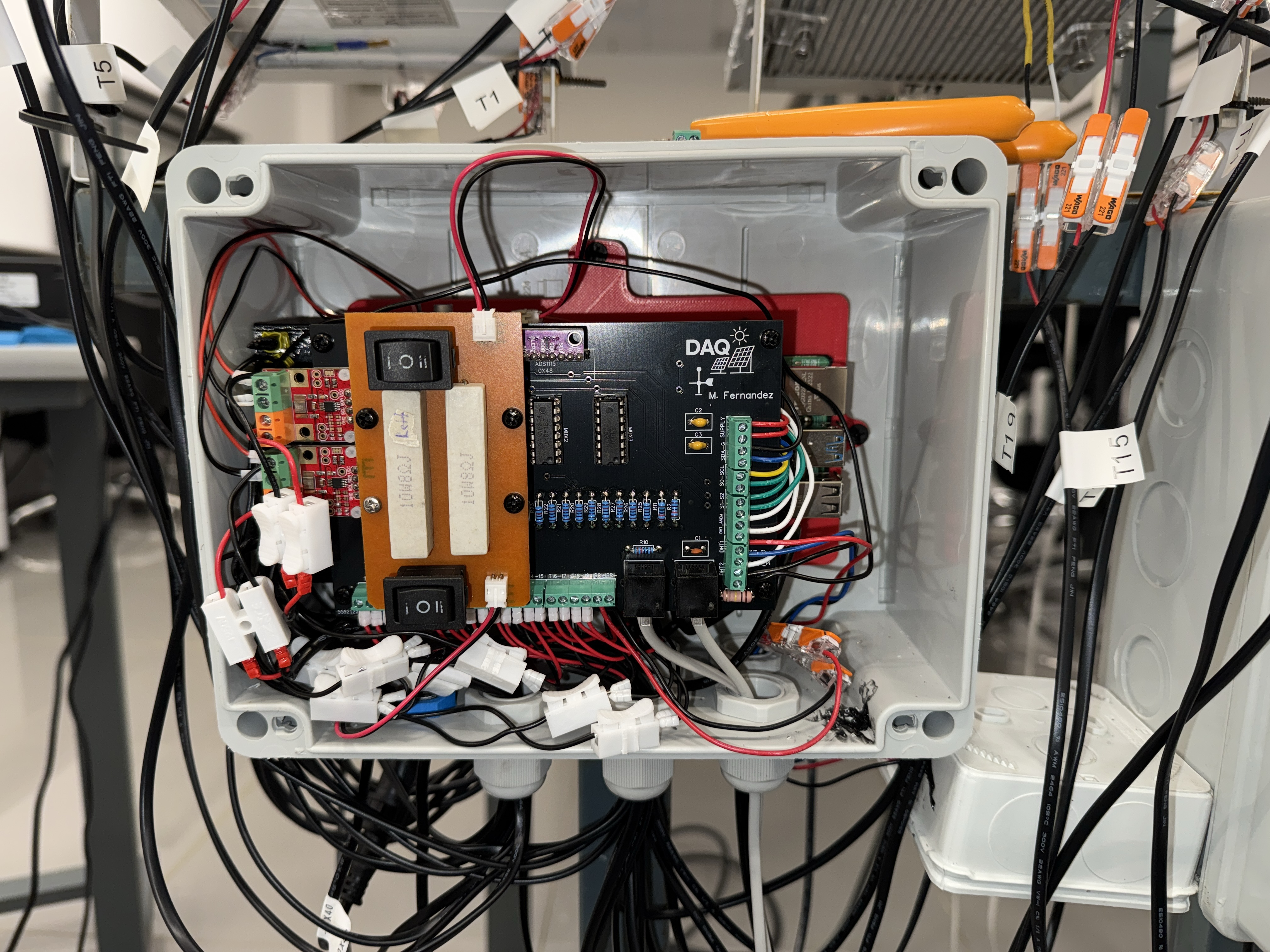}
\caption{Photograph of the hardware used to implement the DAQ.}
\label{fig:Hardware}
\end{figure}

The system was validated by monitoring two \SI{6}{\watt} monocrystalline photovoltaic modules: a control and a treatment modules equipped with a passive heat exchanger (see Fig.\ref{fig:PV_modules}). Both modules operate under identical environmental conditions with independent MPPT controllers, battery banks, and resistive loads. The installation has operated continuously since deployment, capturing data during the defined 5 AM - 6 PM operational window. Figure~\ref{fig:Hardware} shows an image of the hardware´s system developed.

\subsection{Real-Time Data Acquisition and Visualization}

\begin{figure}[htbp]
    \centering
    \includegraphics[width=\columnwidth]{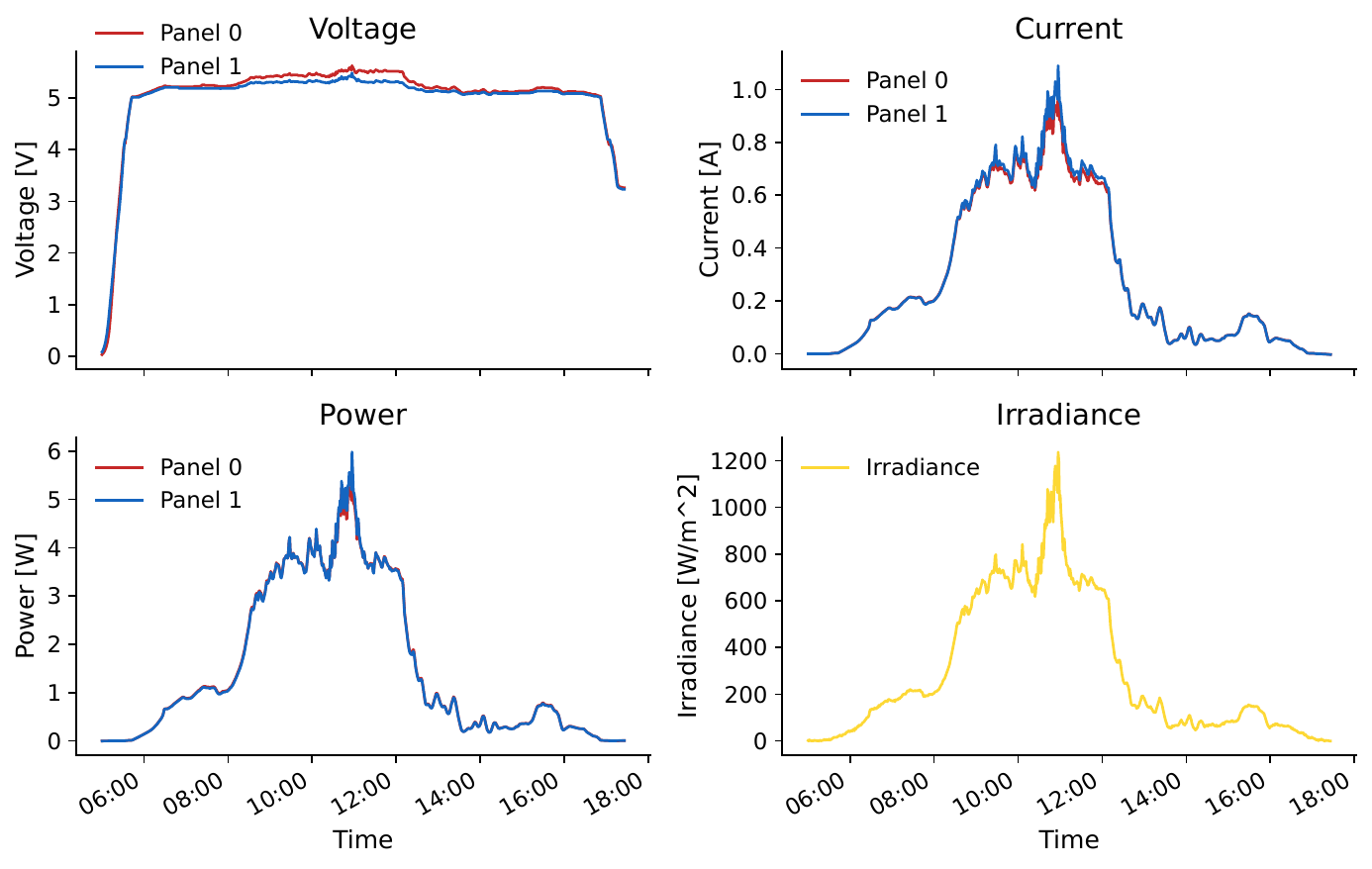}
    \caption{Examples of graphs generated by Grafana from information acquired by the DAQ for time series analysis. In this case, panel 0 and panel 1 operated under the same conditions and both without any external heat exchanger, verifying the accuracy of the measured variables.}
    \label{fig:electrical_comparison}
\end{figure}

Every measurement frame is pushed to InfluxDB and immediately exposed through Grafana dashboards (Fig.~\ref{fig:electrical_comparison}), enabling inspection of live data while the experiment is running. The interface groups channels by function: electrical performance (voltage, current, power, energy), distributed temperatures from the 20-thermistor array, and the complete meteorological stack (irradiance, ambient conditions, wind, precipitation). This organization keeps the comparative PV metrics side-by-side with their environmental context without manual exports or post-processing.

Figure~\ref{fig:electrical_comparison} illustrates how the Grafana dashboard facilitates checking the panel parity, electrical behaviour, and environmental forcing in a single view. Although only the electrical traces are displayed in this excerpt for readability, the live dashboard also exposes meteorological feeds and thermistor readings to contextualize any divergence between the two modules under study.

\subsection{System Reliability and Fault Recovery}

Field deployment demonstrates robust autonomous operation across multiple reliability dimensions:

\textbf{Power Cycle Recovery:} The state recovery mechanism successfully restores operation after unplanned power interruptions. Testing included deliberate power cycles at various times: mid-operation (continuing existing CSV), before operational hours (initializing but remaining idle), and after operational hours (clean state initialization). In all cases, the system correctly evaluated continuation criteria and resumed the appropriate operation mode.

\textbf{Network Resilience:} Data redundancy strategy ensures no data loss during network outages. Local CSV files accumulate measurements independently of network status, with Rclone synchronization resuming automatically upon connection restoration. InfluxDB local deployment eliminates external dependencies for core acquisition functions.

\textbf{Long-term Stability:} Multi-week continuous operation periods (limited only by scheduled maintenance) demonstrate stable performance without memory leaks or resource exhaustion. The multi-threaded architecture maintains a consistent one-minute sampling cadence even during peak CPU usage from concurrent Grafana queries and cloud synchronization operations.

Relative to commercial PV loggers such as Solar-Log~\cite{solar_log_datasheet}, which prioritize aggregated fleet KPIs and assume utility-scale wiring, the demonstrated platform maintains parity in uptime while exposing per-panel thermistor maps, granular meteorological probes, and open data exports. Academic DAQ efforts like~\cite{daq_research} deliver cloud telemetry but lack redundant storage, multi-threaded recovery, or dense thermal instrumentation; the minute-level cadence, dual storage pipeline, and automated reboot flow shown here close those gaps and keep the system operational in harsher field conditions.

\subsection{Data Redundancy and Remote Access}

The dual storage architecture provides complementary access patterns:

\textbf{InfluxDB:} Optimized for time-series queries, aggregations, and dashboard generation. The retention policy provides almost unlimited storage for research applications that require complete historical access. Query performance remains acceptable even with months of minute-resolution data across 20+ parameters.

\textbf{CSV Archives:} Provide portable, tool-agnostic data access suitable for offline analysis. Daily file rotation simplifies dataset management and enables selective data transfer. Google Drive synchronization ensures off-site backup and facilitates data sharing with research collaborators.

Remote monitoring capabilities proved essential during field deployment, enabling troubleshooting and validation without physical site access. SSH connectivity through Pi Connect service provides VPN-free remote terminal access, while the control script interface streamlines common diagnostic operations.

To encourage FAIR-ness of this work, the complete hardware design package, Python acquisition stack, and deployment scripts are available open-source~\cite{zenodo}.

\section{Conclusion}\label{sec:conclusion}

This work presented a modular-designed DAQ system for comparative PV performance analysis, addressing hardware integration, software architecture, and operational deployment considerations. Key achievements include: (1) cost-effective custom PCB integrating 20+ sensor channels through analog multiplexing and I2C aggregation, (2) autonomous operation with state recovery enabling continuous field deployment despite power interruptions, (3) dual storage redundancy eliminating single points of failure for critical research data, and (4) real-time visualization enabling immediate feedback during experiments.

Future enhancements could address: expanded sensor capacity through additional multiplexers, cellular network connectivity for installations lacking WiFi infrastructure, edge computing for on-device analytics, reducing bandwidth requirements, and standardized enclosure designs for accelerated field deployment.

The complete system design files, software source code, and deployment documentation are available online and open source for further research in renewable energy research and industrial installations. This solution shows the capacity of accessible modern technology to support cost-effective, hardware-software platforms for specialized systems analysis. The proposed design serves as a versatile framework that can be readily adapted to other monitoring scenarios with similar high-precision requirements.

\section*{Acknowledgments}

This work was fully funded by the Instituto Tecnológico de Costa Rica (TEC) through the project “Sistema de enfriamiento pasivo para paneles fotovoltaicos mono-faciales”, funding number 1341026.

\end{document}